\documentclass[preprint,12pt]{elsarticle}
\usepackage{amssymb}
\usepackage{booktabs}
\usepackage{dcolumn}
\usepackage{multirow}
\usepackage{adjustbox}
\usepackage{subfig}
\usepackage{graphicx}
\usepackage{natbib} 
\usepackage{booktabs}
\usepackage{lscape}
\usepackage{amsmath}
\usepackage[obeyspaces,hyphens]{url}
\usepackage{xcolor}
\usepackage[countmax]{subfloat}
\biboptions{sort&compress}
\newcommand{\be}{\begin{equation}}
\newcommand{\ee}{\end{equation}}
\newcommand{\beq}{\begin{eqnarray}}
\newcommand{\eeq}{\end{eqnarray}}
\newcommand{\lb}[1]{\label{#1}}

\newcommand{\ssty}{\scriptscriptstyle}
\newcommand{\tsty}{\textstyle}

\newcommand{\dd}{{\mathrm{d}}}
\newcommand{\dl}{d_{\ssty L}}
\newcommand{\dlzero}{d_{\ssty L_0}}
\newcommand{\Vl}{V_{\ssty L}}
\newcommand{\da}{d_{\ssty A}}

\newcommand{\tb}{t_{\ssty B}}
\newcommand{\rb}{r_{\ssty B}}
\newcommand{\gaml}{\gamma_{\ssty L}}

\newcommand{\dobs}{d_{{\tsty {\ssty \rm obs}}}}
\newcommand{\Nobs}{N_{{\tsty {\ssty \rm obs}}}}
\newcommand{\Vobs}{V_{{\tsty {\ssty \rm obs}}}}
\newcommand{\gobs}{\gamma_{{\tsty {\ssty \rm obs}}}}

\newcommand{\etal}{\textit{et al.\ }}
\makeatletter
\newcommand{\oz}{\mathbin{\mathpalette\make@circled{z}}}
\newcommand{\odl}{\mathbin{\mathpalette\make@circled{\dl}}}
\newcommand{\ogams}{\mathbin{\mathpalette\make@circled{\gaml^\ast}}}
\newcommand{\make@circled}[2]{%
  \ooalign{$\m@th#1\smallbigcirc{#1}$\cr\hidewidth$\m@th#1#2$\hidewidth\cr}%
}
\newcommand{\smallbigcirc}[1]{%
  \vcenter{\hbox{\scalebox{1.6}{$\m@th#1\bigcirc$}}}%
}
\makeatother
\usepackage{array}

\journal{}

\begin{document}

\begin{frontmatter}

\title{Galaxy Mergers in a Fractal Cosmology}

\author[1]{Bruno J. Souza}
\address[1]{Physics Institute, Universidade Federal do
Rio de Janeiro, Brazil\fnref{1}}
\ead{brunojsouza@pos.if.ufrj.br}

\author[1]{Osvaldo L.\ Santos-Pereira}
\ead{olsp@if.ufrj.br}

\author[1]{Marcelo B.\ Ribeiro\corref{cor3}%
}
\ead{mbr@if.ufrj.br}

\cortext[cor3]{\it Corresponding author}

\begin{abstract}
This work discusses the influence of galaxy mergers in the evolution of
a parabolic Lema\^{\i}tre-Tolman-Bondi (LTB) cosmology with simultaneous
big bang endowed with two consecutive single fractal galaxy distributions
systems possessing fractal dimension $D$. Based on recent empirical
findings, it is assumed that the resulting galaxy mass from mergers can
be expressed by a redshift dependent decaying power law. The proposed
cosmological model modifies the relativistic fractal number counts
distribution by including a merger rate evolution that estimates the
model's radial density. Numerical solutions for the first order
small-merger-rate approximation (SMRA) are found and the results show
that a fractal galaxy distribution having $D=1.5$ in the range
$0.1<z<1.0$, and $D=0.5$ for $1<z<6$, as suggested by recent empirical
findings, the SMRA allows consistent description of the model for a
merger rate power law exponent up to $q=0.2$ considering a fractal galaxy
distribution starting from the Local Group. Consistent values were also
found up to $q=2.5$ and $z=7$ from a scale smaller than the Local
Supercluster. These results show that galaxy mergers can be successfully
incorporated into the dynamics of a parabolic LTB fractal cosmology.
\end{abstract}

\begin{keyword}
cosmology\sep galaxy mergers\sep fractal galaxy distribution\sep
Lema\^{\i}tre-Tolman-Bondi cosmology\sep
\end{keyword}

\end{frontmatter}


\section{Introduction}

Galaxy mergers occur when two or more galaxies collide into each
other and fuse together to form a new galaxy. The physical effects
of the merger are various, from changing the galaxy morphology, for
instance, from disk galaxies to elliptical ones, to influencing star
formation activity, size growth and the increase of massive objects
at the galaxy centers. Hence, the evolution and formation of galaxies
appear to be intricately linked to their merger history, making galaxy
mergers an important phenomenon to be taken into account in galactic
and cosmological studies \cite{duan2024}.

Despite their importance, galaxy merger processes have not yet been 
considered into the dynamics of relativistic cosmological models. When
mergers are included in cosmological studies their effects are
restricted to the influence in the standard cosmology parameters, not
within the relativistic models themselves \cite{conselice,sure24}.
Relativistic cosmological models where galaxy distributions are viewed
as forming fractal systems are naturally suited to consider mergers
because fractal galaxy distributions have, by construction, overdense
and underdense galactic regions, clusters and voids \cite{pietronero87,
coleman92,sylos98,ribeiro98}, environments where mergers are thought
to increase \cite{sure24}. 

Modeling galaxy mergers in a relativistic fractal cosmology has
challenges of their own. Although it has long been known to be
possible to conciliate fractal galaxy systems with the standard
{Friedmann-Lema\^{\i}tre-Robertson-Walker} (FLRW) cosmological model
\cite{ribeiro92b,emr2001,ribeiro2001a,ribeiro2001b,ribeiro2005,
vinicius2007,juracy2008,iribarrem2012a,iribarrem2012b,gabriela,
Teles2021,Teles2022, Teles2023}, to incorporate galaxy mergers one
has to include this effect into the models' mass distribution in order to
derive a fractal number counts where galaxy mergers can be taken into
account. One way of doing this is to start with the simplest generalization
of the FLRW spacetime, that is, the Lema\^{\i}tre-Tolman-Bondi (LTB)
cosmology, and use its functional freedom to include merger processes.

The LTB model represents a universe dominated by dust with radial
inhomogeneity \cite{Lemaitre34,Tolman69,Bondi47}. It is commonly used
as an alternative to the FLRW model when dealing with observational
data of the large-scale structure of the Universe \cite{Lopes2017}.
The inhomogeneity of these structures can be described by a single
fractal system in which the fractal dimension $D$ represents how
irregular the distribution of galaxies is \cite{ribeiro92a,ribeiro93,
iribarrem2014}.

The LTB cosmological model has three arbitrary functions that depend
on the radial coordinate $r$, and their usual interpretation are the
time past the big bang singularity hypersurface for each observer
situated at particular values of $r$, the total energy and active
gravitational mass \cite[Sec.\ 1.1, and references therein]{nogueira2013}.
The function determining time after the big bang can be eliminated
by either a coordinate transformation or making it a constant by
assuming a simultaneous big bang time, as this is the case in the FLRW
models. Another function can be eliminated by considering the limiting
escape energy case equivalent to the Einstein-de Sitter cosmology. Then
only one arbitrary function remains, situation which renders this
restricted model closest to flat FLRW, but retaining enough functional
freedom that allows a combination of both single fractal galaxy
distributions and galaxy mergers. This article will then explore the
effects of galactic mergers on the parabolic LTB cosmology endowed
with a fractal galaxy system in the Universe.

The fractal galaxy distribution approach to be taken here is the
original one, previously known as hierarchical cosmology as proposed by
Charlier \cite{charlier08,charlier22} and developed by de Vaucouleurs
\cite{vaucouleurs70}, Wertz \cite{wertz70, wertz71} and Pietronero
\cite{pietronero87} in the Newtonian context. This means that in here
number counts will be treated \textit{cumulatively} \cite{gabriela,
Teles2021,Teles2022}. Refs.\ \cite{ribeiro98,ribeiro94} provide
thorough reviews of the original approach to the hierarchical (fractal)
cosmology in the Newtonian framework.

A relativistic version of the original fractal approach was advanced
in Refs.\ \cite{ribeiro92a,ribeiro93,ribeiro94} for the LTB model,
where the past null cone was integrated numerically
\cite{ribeirocomp2002}, as well as for the FLRW cosmology
\cite{ribeiro92b,ribeiro2001b}. Nogueira \cite{nogueira2013} also
discussed a fractal cosmology model in the LTB spacetime, but with
the additional novelty of integrating the LTB null cone by means of
the single past radial null geodesics analytical maneuver \cite[Sec.\ 2.1]
{Mustapha97}, an approach that will also be adopted here. Additional
fractal LTB models were advanced by Cosmai \etal \cite{Cosmai19,
cosmai2023} and Past\'en \& C\'ardenas \cite{pasten2023}, but these
authors introduced the fractal dimension $D$ coupled to the radial
coordinate $r$ rather than any observational distance, an approach
that does not adhere to the original hierarchical (fractal) approach
mentioned above.

Galaxy mergers can be included into the model by adopting the
phenomenologically based galaxy mass evolution advanced by Lopes
and collaborators \cite{Lopes2014,Lopes16}. This methodology depicts
galaxy mass changes due to mergers by means of a semi-empirical
decaying power law having $q$ as the parameter exponent. Then, two
methods are used to study the effects of mergers, the first being an
approximation named as \textit{small-merger-rate approximation} (SMRA)
where only the first order expansion term is used, and whose mergers
having $q>0.2$ at $z>2$ are discarded because they no longer present
fractal power law features. The second method is numeric, whose results
are valid up to $q=2.5$ at $z=7$.

This paper is organized as follows. Sec.\ \ref{ltb} presents the basics
of the LTB cosmological model and the single past radial null geodesic
hypothesis. Sec.\ \ref{galdis} discusses galaxy mergers and the fractal
distribution of matter. Sec.\ \ref{mergers} presents the two methods used
to determine the fractal LTB model with galaxy mergers and its density
distribution. Sec.\ \ref{results} applies the model to a scale starting
at the Local Group of galaxies, as well as starting from a scale a little
less than the Local Supercluster, and discusses the results. Sec.\
\ref{conc} concludes the article.

\section{The Lema\^itre-Tolman-Bondi cosmology}\lb{ltb}

This section briefly reviews some basic results of the LTB cosmological
model, the particular parabolic solution, observables derived from the
LTB geometry and the single past null geodesic analytical maneuver that
allows great analytical simplification of the LTB geodesics.

\subsection{The spacetime geometry}

The LTB metric in comoving spherically symmetric coordinates
$x^{\mu}=(t,r,\theta,\phi)$ may be written as follows \cite{ribeiro92a},
\begin{equation}
\textrm{d}s^2 = \textrm{d}t^2 - \frac{R'^2}{f^2}\textrm{d}r^2-R^2(t,r)
(\textrm{d}\theta^2 + \sin^2\theta \textrm{d}\phi^2). \label{metric}
\end{equation}
The Einstein field equations for the LTB spacetime geometry yield,
\begin{equation}
	{\dot{R}}^2-\frac{F(r)}{2R}={f(r)}^2-1, \label{eqcampo}
\end{equation}
and the local mass density is given by,
\begin{equation}
8\pi\rho(r,t)=\frac{F'}{2R'R^2}. \label{densidadelocal}
\end{equation}
Here the prime means partial derivative with respect to $r$, and the
dot means partial derivative with respect to $t$. The case when
$f^2=1$ simplifies Eq.\ (\ref{eqcampo}) such that its integration
produces the following solution,
\begin{equation}
R(t,r)=\frac{[9F(r)]^{1/3}}{2}[t+\beta(r)]^{2/3},
\label{parabolic1}
\end{equation}
known as the \textit{parabolic LTB model}. Eq.\ (\ref{eqcampo}) has two
more solutions for $f^2>1$ and $f^2<1$, respectively known as the
hyperbolic and elliptic LTB solutions \cite[see][]{ribeiro92a,ribeiro93}.

In the expressions above $F(r)$, $f(r)$ and $\beta(r)$ are arbitrary
functions that come out of the field equations integrations. They are 
interpreted as, respectively, active gravitational mass and total
energy, as can be seen from Eq.\ (\ref{eqcampo}), and time past the big
bang singularity, as viewed in Eq.\ (\ref{parabolic1}).

The parabolic model with simultaneous big bang is obtained by the
following choice,
\be
\beta(r)=\tb,
\lb{parabolicsimultaneous}
\ee
where $\tb$ is a constant value indicating the time past after the
big bang for all observers located at any value of the radial
coordinate. Hence, this choice reduces Eq.\ (\ref{parabolic1}) to
the \textit{parabolic LTB solution with simultaneous big bang} below,
\begin{equation}
R(t,r)=\frac{[9F(r)]^{1/3}}{2}{(t+\tb)}^{2/3},
\label{parabolic}
\end{equation}
which means that in this particular LTB model only the function
$F(r)$ remains arbitrary.

\subsection{Observables}\lb{obs}

The \textit{area distance} $\da(r,t)$ \cite{ellis71} for the LTB model is
given by the expression below \cite{ribeiro92a},
\begin{equation}
(\da)^2=R^2,
\lb{disarea}
\end{equation}
and the \textit{luminosity distance} $\dl(r,t)$ is obtained by means of
the \textit{Etherington reciprocity theorem} \cite{etherington33,ellis71,
ellis2007}, also known as the \textit{distance-duality relation}
\cite{holanda2010,holanda2011,holanda2012}, as follows,
\begin{equation}
\dl=\da(1+z)^2.
\label{dl_da_relation}
\end{equation}

The local number density $n$ may be defined as follows,
\begin{equation}
n(r,t)=\frac{\rho}{\mathcal{M}_g}=\frac{F'}{16\pi \mathcal{M}_g R'R^2},
\lb{numdensi}
\end{equation}
where $\mathcal{M}_g$ is the galaxy rest mass. This means that it is
being assumed here that cosmological sources are mostly composed of
galaxies in the cosmological era of this model.

Now, considering the expression above for $n$, the differential number
counts may be written as below \cite[see][Sec.\ 1.2.2]{ribeiro92a,nogueira2013},
\begin{equation}
\textrm{d}N=4\pi nR^2\frac{R'}{f}\textrm{d}r =
\frac{F'}{4\,\mathcal{M}_g \,f}\textrm{d}r.
\label{dNLTB}
\end{equation}

The luminosity distance allows us to define an \textit{observational
volume},
\begin{equation}
\Vl=\frac{4}{3}\pi {(\dl)}^3,
\label{volumegenerico}
\end{equation}
and the \textit{observer volume density} yields, 
\begin{equation}
\gaml^*=\frac{\mathcal{M}_g N}{\Vl}.
\label{densidadegenerica}
\end{equation}
Notice that although both quantities above were defined with the
luminosity distance $\dl$ they can be defined with any cosmological
distance.

The redshift in the LTB model is given by the following expression
\cite{ribeiro92a},
\begin{equation}
z= \frac{I}{1-I},
\label{z(I)}
\end{equation}
where the function $I$ is determined by the solution of the equation below,
\begin{equation}
\frac{\textrm{d}I}{\textrm{d}r}=\frac{(1-I)}{f}\dot{R'}.
\label{dif I}
\end{equation}
Nogueira \cite[][Sec.\ 1.2.4]{nogueira2013} showed that Eqs.\ (\ref{z(I)})
and (\ref{dif I}) are equivalent to the LTB redshift expression advanced
by Bondi \cite{Bondi47}.

It must be noticed that all expressions above are radially defined and,
therefore, are function of both the radial and time coordinates. Along
the null cone both coordinates are in fact dependent on the null cone
affine parameter $\lambda$, that is, $r=r(\lambda)$ and $t=t(\lambda)$.
However, it is more convenient to carry out null cone integrations
using the radial coordinate $r$ as the independent variable, which has
the effect of turning the affine parameter implicit in all expressions,
that is, $t=t[r(\lambda)]$.

\subsection{The Single Past Radial Null Geodesic}\lb{single}

Mustapha \textit{et al}.\ \cite{Mustapha97} observed that in
cosmology observational events occur on a time scale far greater
than the time lapse of available data on the sky. Therefore, it is
possible to accept that all observational events of cosmological
significance are in practice made along the same single past null
geodesic. This hypothesis allows us to greatly simplify the problem
of null geodesic integration, making it analytically feasible,
otherwise one has to do it numerically \cite{ribeiro93,ribeirocomp2002}.
This analytical approach to null geodesic integration will be very
briefly exposed below.

The LTB  radial null geodesic means assuming that $\textrm{d}s^2=0=
\textrm{d}\theta^2=\textrm{d}\phi^2$ in Eq.\ (\ref{metric}). Then
the \textit{past} radial null geodesic may be written as follows,
\begin{equation}
\frac{\textrm{d}{t}}{\textrm{d}\lambda}=-\frac{R'[t,r]}{f}
\frac{\textrm{d}r}{\textrm{d}\lambda}=-\frac{{R'}}{f}
\frac{\textrm{d}r}{\textrm{d}\lambda}.
\label{Eq2.01}
\end{equation}
A particular past radial null geodesic can be chosen by the relation
below,
\begin{equation}
\frac{\widehat{R'}}{f}=1,
\lb{singlepast}
\end{equation}
where the hat denotes quantities evaluated along this particular 
past radial null geodesic, that is, $\hat{t}=\hat{t}(\hat{r})$ \cite[see
also] [chap.\ 2]{nogueira2013}. Then, Eq.\ (\ref{Eq2.01}) yields,
\begin{equation}
\hat{t}(\hat{r})=C-\hat{r},
\lb{geodesic-1}
\end{equation}
where $C$ is an integration constant. Let us now define the origin
of the single past radial null geodesic as being given by the coordinates
$\hat{t}=0=\hat{r}$, this being our ``here and now.'' Hence, $C=0$
from Eq.\ (\ref{geodesic-1}), which then becomes,
\be
\hat{t}=-\hat{r}.
\lb{geodesic}
\ee
So, the big bang singularity hypersurface has coordinate $\hat{t}=
-\widehat{\tb}$ because at this value one can see from Eq.\
(\ref{parabolic}) that $R=0$ and the metric (\ref{metric}) becomes
singular. From Eq.\ (\ref{geodesic}) this big bang hypersurface also
has the coordinate $\hat{r}=\widehat{\rb}$.

The results above allow the redshift expressions (\ref{z(I)}) and
(\ref{dif I}) along this particular past radial null geodesic to be
readily solved analytically \cite[see][Sec.\ 2.2] {nogueira2013}, yielding, 
\begin{equation}
1+\hat{z}(\hat{r})=\frac{\widehat{\tb}^{\frac{2}{3}}}
{(\widehat{\tb}-\hat{r})^{\frac{2}{3}}}.
\label{z(r)}
\end{equation}

From now on we shall drop the hat to make the notation simpler, but all
calculations from this point on will be along the single past radial
null geodesic defined by Eq.\ (\ref{geodesic}). 

\section{The galaxy distribution}\lb{galdis}

In order to take into account both galaxy mergers and fractal distribution 
in the model being developed here some restrictions on the LTB geometry
must be made. For this purpose the data analysis of the galaxy cosmological
mass function made in Refs.\ \cite{Lopes2014,Lopes16} and the relativistic
version of a self-similar fractal distribution advanced in Ref.\
\cite{ribeiro92a} will allow the expression of galaxy rest mass and number
counts in terms of suitable equations.

\subsection{Galaxy mergers}\lb{mergers2}

Lopes \etal \cite{Lopes2014,Lopes16} analyzed the FDF and UltraVISTA
galaxy surveys and concluded that the data allowed fitting the average
galactic mass to a redshift dependent power law using both the stellar
mass-to-light ratio and galaxy stellar mass function methods. Nevertheless,
Lopes' \cite{Lopes16} analyses showed that the latter methodology
provided much less biased results, yielding a negative power index that
indicates a growth in galaxy mass from high to low redshift values. The
proposed expression is written below
\cite[Secs.\ 5.2, 5.4.2]{Lopes16}
\begin{equation}
\mathcal{M}_g=\mathcal{M}_0(1+z)^{-q}, \label{M_g(z)}
\end{equation}
where $q$ \textit{is the parameter that determines the mass variation
along} $z$, and $\mathcal{M}_0$ is the galaxy mass at $z=0$.

In the context of this work $\mathcal{M}_0$ will be assumed as the lower
cutoff point of the fractal galaxy system, that is, where it actually
begins. This can be either at the scale of the \textit{Local Group} 
of galaxies or somewhat beyond it such as the \textit{Local Supercluster},
but in both cases this means $\mathcal{M}_0 \sim 10^{12} \mathcal{M}_\odot$.
Choosing one or the other depends on the modeling choice of galaxy mergers
(see Sec.\ \ref{results} below).

Regarding the parameter $q$, the empirical analysis of the UltraVISTA
survey yielded the values $q=[(2.08, 2.78, 3.63)\pm0.01]\times10^{-2}$
for the $\Lambda$CDM, parabolic and hyperbolic LTB models, respectively,
whereas the value $q=(0.58\pm0.22)$ was found with the FDF survey data
\cite[pp.\ 86, 93]{Lopes16}. These results indicate that older galaxies
have less mass than new ones, which suggest that the increase in galaxy
mass may be caused by merging.

In summary, Eq.\ (\ref{M_g(z)}) can be used as a semi-empirical expression
for modeling galaxy mergers. 

\subsection{The fractal model}

The original \textit{Pietronero-Wertz hierarchical (fractal) cosmology}
essentially boils down to a key phenomenological hypothesis called
\textit{the number-distance relation}, written as follows \cite{ribeiro94,
ribeiro98},
\be
\Nobs(\dobs)=\sigma(\dobs)^D,
\label{ContagemPietronero}
\ee
where $\dobs$ is the radial observational distance of the observational
volume
\be
\Vobs=\frac{4}{3} \pi {(\dobs)}^3,
\lb{vobs}
\ee
$\Nobs$ is the \textit{cumulative} number counts, $\sigma$ is a
parameter related to the first scale where the fractal system is
defined, and $D$ is the fractal dimension of the structure
\cite[appx.\ B]{Teles2023} \cite[\S 3]{ribeiro94}.
In the original model the \textit{observer number density}
$\widetilde{\gobs^\ast}$ is given by \cite{Teles2022},
\be
\widetilde{\gobs^\ast}=\frac{\Nobs}{\Vobs},
\lb{gobs-ast}
\ee
where the  difference with the observer volume density defined in
Eq.\ (\ref{densidadegenerica}) should be noticed. Substituting Eqs.\
(\ref{ContagemPietronero}) and (\ref{vobs}) into Eq.\ (\ref{gobs-ast})
we obtain the \textit{de Vaucouleurs density power law}
\citep{pietronero87,ribeiro98}, 
\begin{equation}
\widetilde{\gobs^\ast} = \frac{3\sigma}{4\pi}{(\dobs)}^{-(3-D)}.
\lb{gstar3}
\end{equation}
In a relativistic model $\dobs$ must be replaced by some cosmological
distance, not by the radial coordinate $r$ if the model were to become
consistent with the original hierarchical (fractal) cosmology approach.
Hence, here we shall follow Ref.\ \cite{ribeiro92a} and rewrite Eq.\
(\ref{ContagemPietronero}) as follows,
\begin{equation}
N=\sigma (\dl)^D.
\label{ContagemFractal}
\end{equation}
By following this analytical route the number counts become dependent on
the luminosity distance $\dl$, which is an observable distance in
relativistic models and can be easily related to other cosmological
distances \cite{ellis2007} by means of Eq.\ (\ref{dl_da_relation}).

\section{Galaxy mergers in a fractal parabolic LTB}\lb{mergers}

The next step in our modeling is to find the expression for number
counts, which means integrating Eq.\ (\ref{dNLTB}) for further use
in Eq.\ (\ref{densidadegenerica}). Considering the single past null
cone discussed above that leads to Eq.\ (\ref{z(r)}), and that Eq.\
(\ref{M_g(z)}) already embodies the galaxy mergers hypothesis, then
in the specific modeling under consideration here Eq.\ (\ref{dNLTB})
may be rewritten as below, 
\begin{equation}
N(r)=\frac{(\tb)^{2q/3}}{4\mathcal{M}_0}\int_{0}^{r}
\frac{F'\textrm{d}r}{f(\tb-r)^{2q/3}},
\label{Ncomfusao}
\end{equation}
where
\be
f(r)=1,
\lb{f=1}
\ee
according to the choice of the parabolic LTB model and to be
consistent with the past null cone defined by Eq.\ (\ref{Eq2.01}).
It is clear from Eq.\ (\ref{Ncomfusao}) that the task now is to adjust
the remaining arbitrary function $F(r)$ and its derivative $F'(r)$
such that they together describe both the fractal galaxy system
as well as galaxy mergers. 

In order to solve the integral in the expression (\ref{Ncomfusao}) we
first need to determine $F'(r)$ in Eq.\ (\ref{dNLTB}). This can be
done in two ways:
\begin{enumerate}[(1)]
\item by Taylor expanding Eq.\ (\ref{M_g(z)}) to find a SMRA for $F(r)$;
\item by finding $F(r)$ through numerical means.
\end{enumerate}

\subsection{The small-merger-rate approximation (SMRA)}\lb{smra}

Let us first consider galaxy mergers at low redshift and low values of
the parameter $q$. The Taylor series of Eq.\  (\ref{M_g(z)}) may be
written as below,
\begin{equation}
{\mathcal{M}}_g= {\mathcal{M}}_0(1+z)^{-q}={\mathcal{M}}_0
-{\mathcal{M}}_0qz+{\mathcal{M}}_0\left(\frac{q^2}{2}+
\frac{q}{2}\right)z^2+...\,.
\lb{taylorMg}
\end{equation}
For small values of $z$ one may truncate the series up to the first
order term of ${\mathcal{M}}_g(z)$, that is,
\begin{equation}
{\mathcal{M}}_g\approx {\mathcal{M}}_0(1-qz),
\label{MGaproximado}
\end{equation}
which allows Eq.\ (\ref{dNLTB}) to be rewritten as follows,
\begin{equation}
N\approx \frac{1}{4{\mathcal{M}}_0}\int_{0}^{r}\frac{F'}{(1-qz)}\textrm{d}r.
\lb{dNLTB2}
\end{equation}

The SMRA is the condition where $qz\ll1$, which means either very small
values of $z$, or $q$, or both. In addition, in this approximation one
may assume that within the variation range of the radial coordinate $r$
in Eq.\ (\ref{dNLTB2}) the term $(1-qz)$ changes very little and,
therefore, may be taken out of the integral and then $N$ may be rewritten
as follows,
\begin{equation}
N\approx \frac{F(r)}{4{\mathcal{M}}_0(1-qz)}.
\lb{dNLTB3}
\end{equation}
Further algebraic manipulations are now possible. Considering Eq.\
(\ref{z(r)}) the number counts turn out to be,
\begin{equation}
N(r)\approx F  \Biggl( 4{\mathcal{M}}_0\bigg\{
1-q\bigg[\displaystyle\frac{{\tb}^{2/3}}
{(\tb-r)^{2/3}}-1\bigg]\bigg\} \Biggr)^{-1}.
\label{Napprox}
\end{equation}
In a fractal LTB model Eqs.\ (\ref{ContagemFractal}) and (\ref{Napprox})
describe the same number counts, hence, 
\begin{equation}
F\Bigg( 4{\mathcal{M}}_0\bigg\{1-q\bigg[\displaystyle\frac{{\tb}^{2/3}}
{(\tb-r)^{2/3}}-1 \bigg]\bigg\} \Bigg)^{-1}\mkern-10mu=\sigma(\dl)^D.
\lb{FdL1}
\end{equation} 
Considering Eqs.\ (\ref{parabolic}), (\ref{disarea}),
(\ref{dl_da_relation}), (\ref{geodesic}) and (\ref{z(r)}) one can
express $\dl$ in terms of $r$, and the equation above becomes,
\begin{equation}
F\Bigg( 4{\mathcal{M}}_0\bigg\{1-q\bigg[\displaystyle\frac{{\tb}^{2/3}}
{(\tb-r)^{2/3}}-1\bigg]\bigg\} \Bigg)^{-1}\mkern-10mu=\sigma\bigg[\frac{1}
{2}(9F)^{1/3}(\tb-r)^{2/3}
\frac{{\tb}^{4/3}}{(\tb-r)^{4/3}}\bigg]^D,
\end{equation}
which allows $F(r)$ to be isolated and written in terms of $r$,  
\begin{equation}
\begin{split}		
F(r)=&(4\sigma {\mathcal{M}}_0)^{3/(3-D)}\Bigg\{1-q\bigg[
\frac{{\tb}^{2/3}}{(\tb-r)^{2/3}}-1\bigg]\Bigg\}^{3/(3-D)}\\
&\times\Bigg[\frac{9^{1/3}}{2}\frac{{\tb}^{4/3}}
{(\tb-r)^{2/3}}\Bigg]^{3D/(3-D)}.
\label{Fdersemnorm}
\end{split}
\end{equation}
The regularity conditions for the LTB geometry state that near $r=0$
the metric (\ref{metric}) must be Euclidean \cite[Sec.\ 3.1]{ribeiro93},
then to be regular we must have $F(0)=0$. To fulfill this condition
in the expression above a term must be subtracted, yielding,
\begin{equation}
\begin{split}
	F(r)=&\;(4\sigma {\mathcal{M}}_0)^{3/(3-D)}\\
      &\times\Bigg(\bigg\{1-q\bigg[\frac{{\tb}^{2/3}}{(\tb-r)^{2/3}}
     -1\bigg]\bigg\}^{3/(3-D)}\bigg[\frac{9^{1/3}}{2}\frac{{\tb}^{4/3}}
     {(\tb-r)^{2/3}}\bigg]^{3D/(3-D)}\\
     &-\bigg[\frac{9^{1/3}}{2}{\tb}^{2/3}\bigg]^{3D/(3-D)}\Bigg),
\label{Fdernormalizado}
\end{split}
\end{equation}
which may be rewritten as,
\begin{equation}
\begin{split}
&F(r)=\;\left[2^{(2-D)/D}\,\sigma^{1/D}\,{{\mathcal{M}}_0}^{1/D}\,9^{1/3}
\right]^{3D/(3-D)}\Big[{\tb}^{2D/(3-D)}\Big]\\
     &\times\Bigg(\bigg\{1-q\bigg[\frac{{\tb}^{2/3}}{(\tb-r)^{2/3}}
     -1\bigg]\bigg\}^{3/(3-D)} 
     {\bigg[\frac{{\tb}^{2/3}}
     {(\tb-r)^{2/3}}\bigg]}^{3D/(3-D)}-1\bigg]\Bigg).
\label{Fdernormalizado2}
\end{split}
\end{equation}

The expression for the luminosity distance $\dl$ (Eq.\
\ref{dl_da_relation}) can now be obtained in terms of the radial
coordinate $r$ by considering Eqs.\ (\ref{parabolic}), (\ref{disarea}),
(\ref{z(r)}), (\ref{Fdernormalizado2}). Rearranging the exponents and
simplifying terms $\dl$ yields ,
\begin{equation}
\begin{split}
	&\dl(r)=\;\;\bigg[\frac{9\,\sigma\,{\mathcal{M}}_0\,(\tb)^2}
	{2}\bigg]^{1/(3-D)}\bigg[\frac{\tb}{(\tb-r)}\bigg]^{2/3}\\
	&\times \Bigg(\bigg\{1-q\bigg[\frac{{\tb}^{2/3}}{(\tb-r)^{2/3}}
	-1\bigg]\bigg\}^{3/(3-D)}\bigg[\frac{{\tb}^{2/3}}
	{(\tb-r)^{2/3}}\bigg]^{3D/(3-D)}-1\Bigg)^{1/3}.
\label{dldernormalizado}
\end{split}
\end{equation}

Bringing together Eqs.\ (\ref{volumegenerico}), (\ref{ContagemFractal})
and (\ref{MGaproximado}) the observer volume density $\gaml^*$ (Eq.\
\ref{densidadegenerica}) produces the expression below,
\be
\gaml^\ast=\frac{3\sigma\mathcal{M}_0}{4\pi}(1-qz){\dl}^{-(3-D)},
\lb{densidadedernormalizada1}
\ee
whose rewritten form, which takes into account Eqs.\ (\ref{z(r)})
and (\ref{dldernormalizado}), becomes,
\begin{equation}
\begin{split}
&\gaml^*(r)=\;\frac{1}{6\pi \tb^2}\bigg\{1-q\bigg[\frac{\tb^{2/3}}
{(\tb-r)^{2/3}}-1\bigg]\bigg\}\left[\frac{\tb}{(\tb-r)}
\right]^{2(D-3)/3}\\
&\!\!\!\!\!\!\times\Bigg(\bigg\{1-q\bigg[\frac{\tb^{2/3}}
{(\tb-r)^{2/3}}-1\bigg]\bigg\}^{3/(3-D)}\bigg[\frac{\tb^{2/3}}
{(\tb-r)^{2/3}}\bigg]^{3D/(3-D)}-1\Bigg)^{(D-3)/3}.
\label{densidadedernormalizada}
\end{split}
\end{equation}

Notice that both $\dl$ and $\gaml^\ast$ in Eqs.\ (\ref{dldernormalizado})
and (\ref{densidadedernormalizada}) are parameterized by the radial
coordinate $r$, which has a maximum value that corresponds to the big
bang singularity hypersurface (see Sec.\ \ref{results} below). Beyond this
maximum both expressions are no longer valid. 

The de Vaucouleurs density power law (\ref{gstar3}) shows that a
fractal galaxy distribution whose dimension is smaller than 3 will
express itself as a decaying straight line in the $\gaml^\ast$ vs.\
$\dl$ log-log plot. The smaller the values of D the steeper the
power law decay, which means more gaps in the galaxy distribution,
that is, galaxy clusters become more and more apart from each other
with fewer and fewer galaxy groups in between. Hence, the
characterization of a fractal system means log-log plotting
$\gaml^\ast$ vs.\ $\dl$ with Eqs.\ (\ref{dldernormalizado}) and
(\ref{densidadedernormalizada}) and observing the appearance of a
straight line with a negative slope, situation which then makes it
possible to evaluate the validity of the approximation.

Notice that in this scenario the galaxy merger rate $q$ becomes an
intrinsic feature of the model, and due to the high nonlinearity of both
expressions (\ref{dldernormalizado}) and (\ref{densidadedernormalizada})
its variation might alter the behavior of both quantities even for a
fixed $D$. Hence, some values of $q$ may nullify the approximation
(see Sec.\ \ref{results} below).

\subsection{The numerical method}\lb{num}

The second method that will be used to determine the function $F(r)$
in the model scenario proposed here is numeric root finding. The
starting point of this approach is the realization that Eqs.\
(\ref{ContagemFractal}) and (\ref{Ncomfusao}) for the number counts
describe the same quantity and, therefore, they can be identified with
each other as below,
\begin{equation}
\frac{(\tb)^{2q/3}}{4\mathcal{M}_0}\int_{0}^{r}\frac{F'\dd r}
{f(\tb-r)^{2q/3}}=\sigma(\dl)^D.
\lb{numeric1}
\end{equation}
Considering Eqs.\ (\ref{parabolic}), (\ref{disarea}),
(\ref{dl_da_relation}), (\ref{geodesic}), (\ref{z(r)}) and (\ref{f=1}),
$\dl$ can be written as follows,
\be
\dl= \frac{1}{2}(9F)^{1/3}
\frac{\tb^{4/3}}{(\tb-r)^{2/3}}.
\lb{numeric3}
\ee
Substituting Eq.\ (\ref{numeric3}) into Eq.\ (\ref{numeric1}), 
differentiating the resulting expression in terms of $r$ and
isolating $F'$ we end up with the following result, 
\begin{equation}
F'=\frac{-2A}{AF^{-1}(\tb-r)-F^{-D/3}(\tb-r)^{(3+2D-2q)/3}},
\label{F'comAsimplificada}
\end{equation}
where $A$ brings together all parameters: 
\begin{equation}
A = 2^{(2-D)}\,3^{(2D-3)/3}\,\sigma\,\mathcal{M}_0 \,D\,
(\tb)^{(4D-2q)/3}.
\lb{defA}
\end{equation}

The single past radial null geodesic  hypothesis (\ref{singlepast})
allows us to determine another expression for $F'$. Using the
derivative of Eq.\ (\ref{parabolic1}) \cite[see][eq.\ 7]{ribeiro92a}
and considering the parabolic LTB model defined by Eqs.\
(\ref{parabolicsimultaneous}) and (\ref{f=1}), as well as Eq.\
(\ref{geodesic}), we may write the following expression,
\begin{equation}
\frac{R'}{1}=1=\frac{1}{3}\left[\frac{9F}{(\tb-r)}\right]^{1/3}
\left[\frac{(\tb-r)F'}{2F}\right],
\label{eqdeF'parabolic}
\end{equation} 
which can be further simplified to,
\begin{equation}
F'=2(3F^2)^{1/3}(\tb-r)^{-2/3}.
\label{derivadadeFparabolic}
\end{equation}
Eqs.\ (\ref{F'comAsimplificada}) and (\ref{derivadadeFparabolic})
describe the same function $F'$ and, therefore, they can be equalized,
yielding,
\begin{equation}
-3^{-1/3}A=AF^{-1/3}(\tb-r)^{1/3}-(F)^{(2-D)/3}(\tb-r)^{(1+2D-2q)/3}.
\lb{F1}
\end{equation}
This expression allows the determination of $F$ in terms of $r$ as an
independent variable by a numeric root finding algorithm. Alternatively,
Eq.\ (\ref{z(r)}) allows changing the independent variable from $r$ to
$z$ in Eq.\ (\ref{F1}). The resulting expression is written below:
\begin{equation}
AF^{-1}(1+z)^{-1/2}\,{\tb}^{1/3}-F^{-D/3}(1+z)^{(2q-2D-1)/2}
\,{\tb}^{(1+2D-2q)/3}+3^{-1/3}AF^{-2/3}=0.
\lb{F2}
\end{equation}

The numerical task now becomes determining a list of $F$ by
fixing different values of $r$ or $z$ in the range from $z=0.1$ to
$z=6$ and finding the root of either Eq.\ (\ref{F1}) or (\ref{F2})
for each value of either $r$ or $z$. If $F$ is being determined by
Eq.\ (\ref{F2}), by writing Eq.\ (\ref{parabolic}) in terms of $z$
using Eqs.\ (\ref{geodesic}), (\ref{z(r)}) and (\ref{f=1}) we end
up with the expression below,
\be
R(z)=\frac{1}{2}{[9F(z)]}^{1/3}\left[\frac{{\tb}^{2/3}}{(1+z)}
\right].
\lb{Rz}
\ee
In this manner the numerical results of $F$ and $R$ allow a list of
$\dl$ to be compiled by means of Eqs.\ (\ref{disarea}) and
(\ref{dl_da_relation}). Similarly numerical lists of $\Vl$,
$\gaml^\ast$, $z$, $M_g$, and $N$ can also be obtained by means of,
respectively, Eqs.\ (\ref{volumegenerico}), (\ref{densidadegenerica}),
(\ref{z(r)}), (\ref{M_g(z)}) and (\ref{ContagemFractal}). The table
below shows schematically how the numerical lists of all quantities
are produced, where the key quantities are inside circles, and the
order of their numerical determination, from left to right.
\begin{center}
\setlength\extrarowheight{4pt}
\begin{tabular}{|m{0.2cm}|m{0.6cm}|m{0.3cm}|m{0.3cm}|m{0.6cm}|m{0.3cm}
	|m{0.5cm}|m{0.3cm}|m{0.6cm}|}
\hline
$r$ & $\oz$ & $F$ & $R$ & $\odl$ & $N$ & $M_g$ & $\Vl$ & $\ogams$ \\
\hspace{-1mm} $\vdots$ & \hspace{1mm} $\vdots $ & \hspace{0mm} $\vdots$
& \hspace{0mm} $\vdots$ &
\hspace{1mm} $\vdots$ & \hspace{0mm} $\vdots$ & \hspace{0.5mm} $\vdots$ &
\hspace{0.2mm} $\vdots$ & \hspace{1mm} $\vdots$
\\ 
\hline
\end{tabular}
\end{center}

\section{Results}\lb{results}

To proceed with the numerical evaluation of the model in well known 
astronomical objects as examples, we need first to establish the 
respective units and parameter values.

Let us start with the units by choosing $3.26\times10^9$ years as the
\textit{time unit}, $2.09\times 10^{22} \mathcal{M}_\odot$ as the
\textit{mass unit}, and $10^3$ Mpc as the \textit{length unit}. These
choices yield $c=G=1$, and $\tb=4.3$ \textit{time units} \cite[Fig.\
2.2]{nogueira2013}.

Regarding the parameters, according to Eq.\ (\ref{ContagemFractal})
$\sigma$ has dimension of $\left[({length})^{-D} \right]$, and
since this parameter defines the lowest scale where the fractal galaxy
distribution starts, let us rewrite Eq.\ (\ref{ContagemFractal})
as follows,
\be
\sigma=N_0 {(\dlzero)}^{-D},
\label{ContagemFractal2}
\ee
where $N_0$ is the number of galaxies at the lower fractal bound and
$\dlzero$ its respective length. For the purposes of this work we may
assume that the lower cutoff of the fractal galaxy distribution, which 
defines $\sigma$ through the expression above, may be given by the
two possibilities chosen as examples for the model's applicability, as
follows.
\begin{enumerate}[(i)]
\item  The fractal galaxy system starts at the level of the Local Group.
       In this case $N_0 \approx 55$ and $\dlzero \approx 3\;\mbox{Mpc}=
       3\times 10^{-3}$ \textit{length units}. Assuming the total mass
       of the Local Group as being $1.3 \times10^{12} \mathcal{M}_\odot$,
       then $\mathcal{M}_0=0.63 \times10^{-10}$ \textit{mass units}.
\item  The fractal galaxy system starts at about 100 Mpc with a
       thousand galaxies. This is much less than the known parameters
       for the Local Supercluster, but more than the Local Group on both
       accounts.  Hence, for this case $N_0 \approx 10^3$, $\dlzero=0.1$
       \textit{length units}, and the total galactic mass estimated as
       being about $2\times10^{12} \mathcal{M}_\odot$ means
       $\mathcal{M}_0=10^{-10}$ \textit{mass units}.
\end{enumerate}

Specific fractal dimensions were chosen according to the empirical 
results indicated in Ref.\ \cite{Teles2021}, these being $D=1.5$ for
the redshift range $0.1<z<1.0$, and $D=0.5$ for $1<z<6$. These are
approximate general values taken from the values calculated from
actual redshift survey data, and as we shall see below the simulations
can extrapolate these redshift ranges in order to determine up to where
both models presented in Secs.\ \ref{smra} and \ref{num} remain valid.
As for the lower cutoffs of the fractal systems, according to Eq.\
(\ref{ContagemFractal2}) we conclude that $\sigma=3.35\times10^5\;
(length\;units)^{-1.5}=10.6\;{\mbox{Mpc}}^{-1.5}$ for $D=1.5$ in case
(i) above, and $\sigma=3.16 \times10^3\;(length\;units)^{-0.5}=100\;
{\mbox{Mpc}}^{-0.5}$ for $D=0.5$ in case (ii).

The scales (i) and (ii) were chosen as to be empirically reasonable
examples for the purposes of this paper, but clearly other scales are
also possible with different values for $\sigma$. Even micromodeling is
possible using the methods above, that is, for very small regions having
short redshift ranges, where, say, empirical data of specific galaxy
clusters are well determined. Successive regions with different values
of $q$ and $D$ are also possible to model. In addition, multiple values
of the galaxy merger rate $q$ can be used to determine the limits of
validity for both methods described in Secs.\ \ref{smra} and \ref{num}. 

\begin{figure}[h]
	\hspace{-0.9cm}
	\includegraphics[scale=0.525]{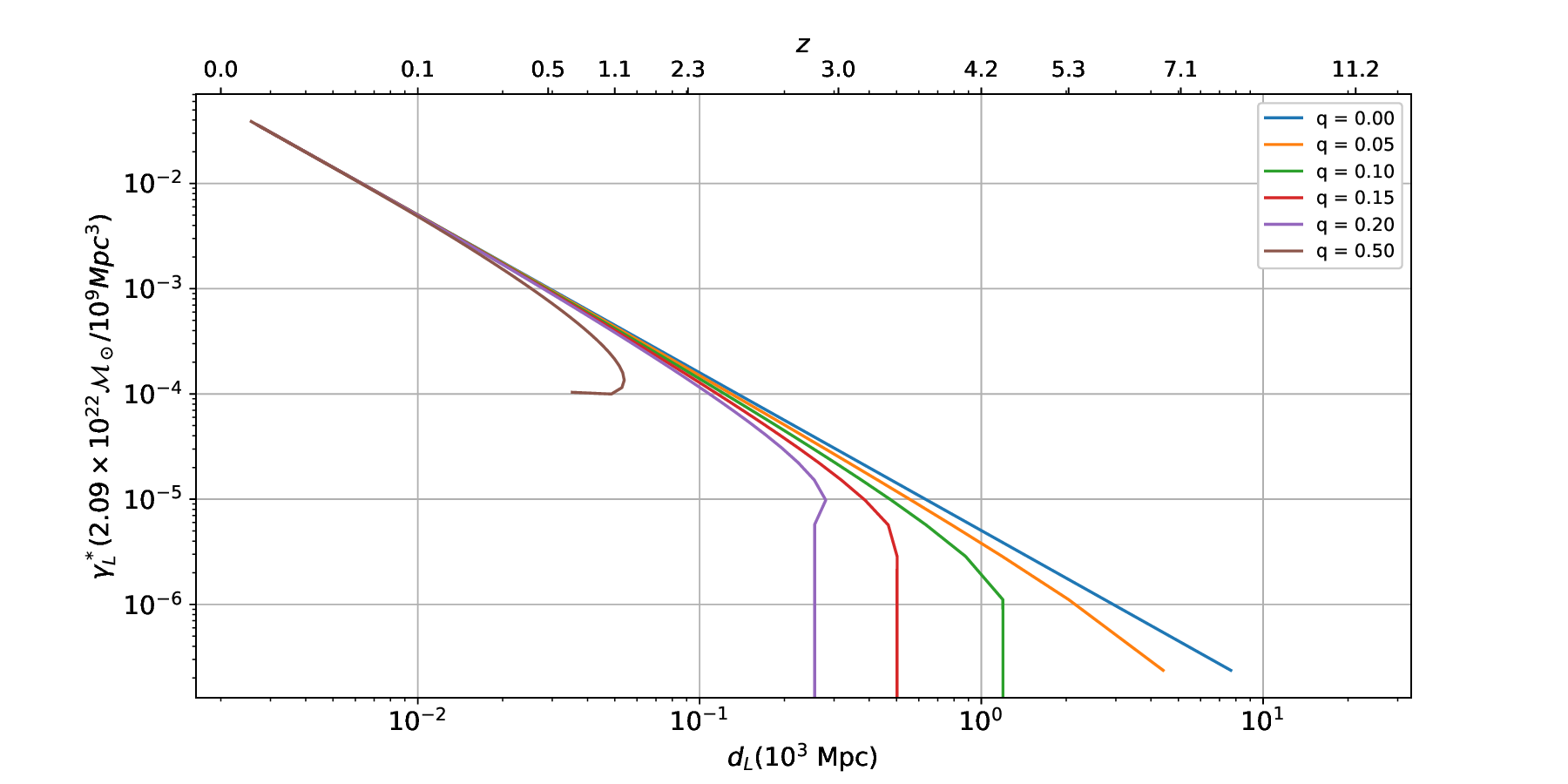}
	\caption{SMRA results for the method discussed in Sec.\ \ref{smra}
	and obtained by means of Eqs.\ (\ref{dldernormalizado}) and
	(\ref{densidadedernormalizada}) using $D=1.5$ whose fractal system
	tarts at the scale of the Local Group. The results of these closed
	functions were extrapolated well beyond $z=1$ to show that the SMRA
	breaks down for the merger rate $q>0.2$ only at $z\sim 2.5$. Results
	for $q=0.5$ remain, however, valid up to $z=0.5$. For $q=0.05$ the
	linearity that characterizes a fractal galaxy distribution remains
        valid at $z=6$.}
	\label{smra.eps}
\end{figure}

Fig.\ \ref{smra.eps} shows a graph of $\gaml^*$ vs.\ $\dl$ for the SMRA,
discussed in Sec.\ \ref{smra}, with multiple plots having different
values for $q$ for a fixed $D=1.5$. It is clear that this approximation
breaks down for larger values of $q$ and $z$, as expected. Most results
show compliance with the fractal model because in it the luminosity
distance is related to the density through a power law with negative
slope represented by a straight line in a log-log plot, as indicated in
Eq.\ (\ref{gstar3}). When the solutions diverge from straight lines this
means that the model stops working and loses its meaning, hence this
methodology allows us to determine the limits of validity of each
submodel. Bearing this in mind, notice that even the high value of
$q=0.5$ allows compliance with the basic premises of the model up to
$z=0.5$. And for $q=0.05$ the SMRA remains valid up to $z=6$.

\begin{figure}[h]
	\hspace{-0.8cm}
	\includegraphics[scale=0.525]{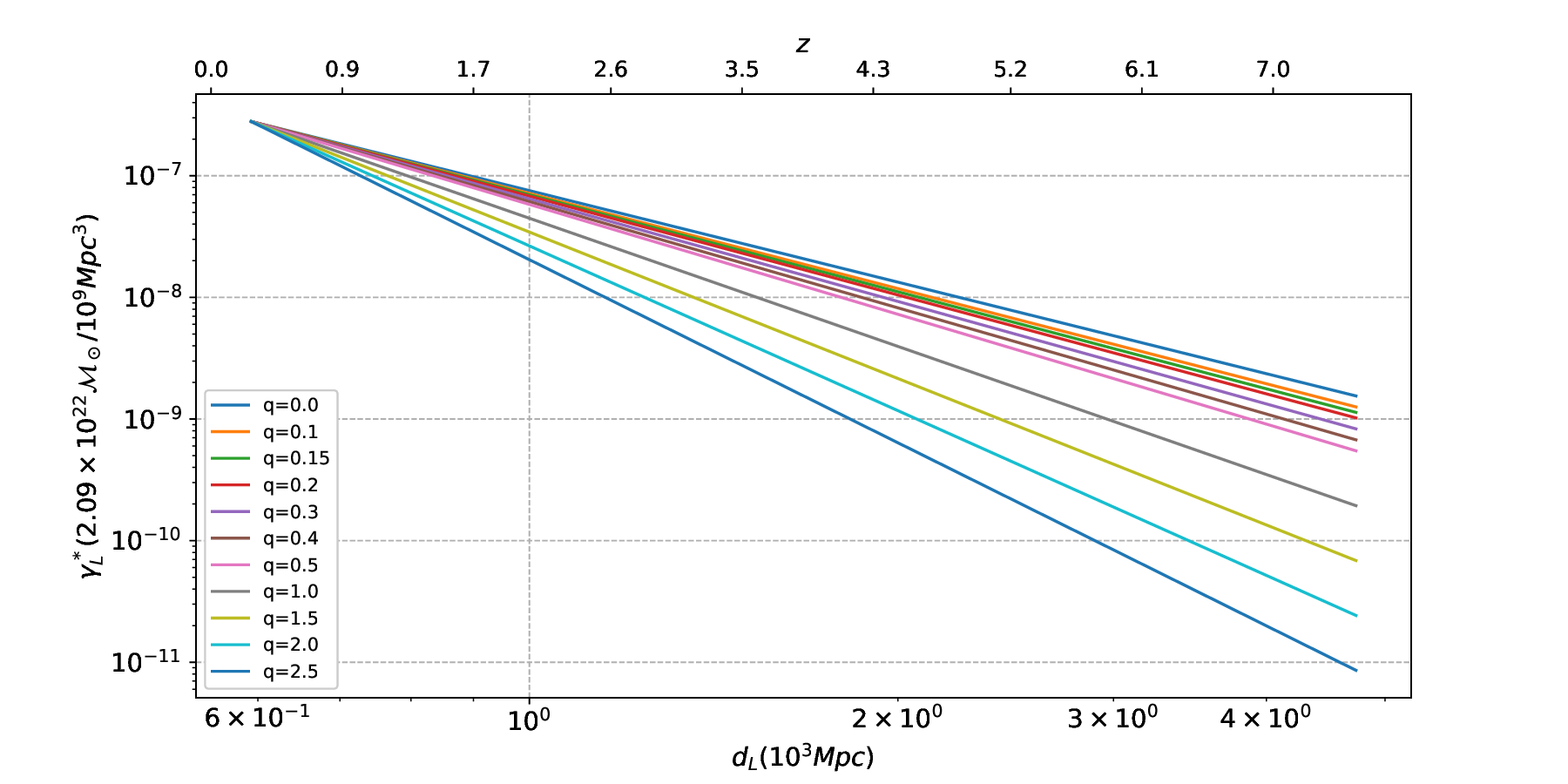}
	\caption{Results obtained through the numerical method discussed
	in Sec.\ \ref{num} and with fixed fractal dimension $D=0.5$ for
	a fractal galaxy system starting the scale a little less than the
	Local Supercluster. Eq.\ (\ref{F1}) was used to obtain the results.
	The plots were extrapolated to $z<1$ and $z\gtrsim7$ and show
	remarkable compliance with the fractal behavior indicated by the
	de Vaucouleurs power law (\ref{gstar3}). The change in the slope
	for higher values of $q$ is consequence of the nonlinearity of
	the problem. However, successful integration required careful
	manipulation of the model's parameters.}
	\label{numeric.eps}
\end{figure}

Fig.\ \ref{numeric.eps} shows the results for the application of the
numerical method discussed in Sec.\ \ref{num}. Plots of $\gaml^*$ vs.\
$\dl$ show linearity from $z\sim0$ to $z\gtrsim7$ and beyond with the
merger rate range $0\le q\le 2.5$. The higher the value of $q$, the
steeper the slope even with fixed $D=0.5$ in all results, a feature that
is certainly a result of the high nonlinearity of Eq.\ (\ref{F1}) used
to compute the results. Eq.\ (\ref{F2}) was not used because it led to
catastrophic loss of significant digits during the root finding
algorithm. It must also be noticed, however, that the numerical method
only worked when the lower cutoff of the fractal galaxy system was
pushed to a higher scale than the Local Group, in fact closer to the
Local Supercluster, since at smaller scales the parameters of the model
led to an effectively vanishing numeric values for $A$ in the expression
(\ref{defA}), which led to the disappearance of terms having $A$ in Eq.\
(\ref{F1}) and its effective break down.

These results show that the numerical method is more reliable than the
SMRA, since it works on a wider range of both redshift and merger rate
parameter $q$. Even so, the SMRA worked more successfully than expected
for $z>1$, with low deviations from the power law despite being  based
on a first order Taylor truncation. And it has the advantage of having
closed functions that do not require further numeric manipulations.

\section{Conclusions}\lb{conc}

In this paper we studied the effect of galaxy mergers in a parabolic
Lema\^{\i}tre-Tolman-Bondi (LTB) cosmological model considering the
single past radial null geodesic hypothesis and in which the
inhomogeneity of the universe is described by a fractal system with
single fractal dimension $D$. Galactic masses were assumed to change
with the redshift through mergers by means of an empirically based
decaying power law index $q$, called merger rate, that indicates growth
in the galactic masses from high to low redshift values.

The parabolic LTB cosmology is characterized by a simultaneous big bang
singularity hypersurface, restriction which leaves only one of its 
arbitrary functions still undetermined, namely $F(r)$, which is
interpreted as the model's gravitational mass. This arbitrariness in
$F(r)$ was then used for depicting both the fractal galaxy distribution
characterized by the single fractal dimension $D$, and the galaxy merger
rate $q$. Here $r$ is the radial coordinate which serves as the
parametric variable of the LTB model. The other parameters of the model
are the lower cutoff $\sigma$ of the fractal system, the overall mass
parameter $\mathcal{M}_0$ at the fractal system's lower bound, and the
time after the big bang $\tb$. 

The determination of the function $F(r)$ such that it indeed describes
the above mentioned features was carried out in two manners. The first
called as \textit{small-merger-rate approximation} (SMRA) requires that
the empirical power law term describing the mass variation be truncated
to first order series expansion for small values of the term $qz$. This
allows writing closed formulas for the luminosity distance $\dl$ and
the observer volume density $\gaml^*$ in terms of the radial coordinate
$r$, and this also allows the determination of the other quantity of
interest, the redshift $z(r)$. The second method does not require any
approximation, but $F(r)$ has to be obtained by numerical means using a
root finding algorithm.

The model was applied to the scale of the Local Group of Galaxies for
the SMRA, and at a scale smaller than the Local Supercluster in the
case of the second method requiring a root finding algorithm to
determine $F$. Fractal dimensions equal to $D=1.5$ and $D=0.5$ were
respectively used for the SMRA and the numerical method. Both of these
values for $D$ were suggested by recent empirical findings of fractal
analysis of galaxy redshift surveys. Plots of $\gaml^*$ vs.\ $\dl$ were
obtained in both cases, and according to the premises of the model these
functions must follow the de Vaucouleurs density power law, therefore
decaying as straight lines in a log-log plot. Deviations from straight
lines mean the breakdown of the model.

The SMRA results considering a fractal galaxy distribution starting
at the scale of the Local Group show that the approximation breaks down
for $q>0.2$ at $z\sim2.5$. For $q=0.05$ the straight line that
characterizes a fractal system remains valid up to $z=6$. These are
remarkable results considering the simplicity of the approximation. But,
for $q=0.5$ the SMRA remains valid only up to $z=0.5$. On the other
hand, the second numerical method, whose starting fractal scale was
somewhat smaller than the Local Supercluster, showed remarkable
stability, being valid in the redshift range $0.1\lesssim z \lesssim 7$
and merger rate $0.1<q<2.5$ and possibly beyond these ranges.

It should be noted that although the model presented here
allows for describing the galaxy fractal behavior and galaxy mergers
at very large scales, such extrapolations must be viewed with caution.
At present, observational results indicate that the galaxy distribution
can be well described as a fractal system of dimension $D\lesssim 2$ up to 
$\sim 100$ Mpc \cite{einasto2020}, even though according to some authors 
\cite{Teles2022,melia2023} the maximum scale could be bigger. Whatever the 
case, one should bear the fact that this model is critically dependent on 
the observational range validity of both Eqs.\ (\ref{M_g(z)}) and 
(\ref{ContagemPietronero}), these being the two key ingredients of the 
model. Hence, very large scale redshift extrapolations currently lack 
observational support.

As final words, the results obtained in this paper show that the novel
modeling approach proposed here is capable of successfully accounting
for the empirical data of the galaxy system, both in terms
of its distribution and galaxy mass variations up to the present
observational ranges. In particular, it seems possible to state that at
nearby ranges the simplest approach taken here, the SMRA, is well capable
of describing observational data on galaxy distribution and merger
evolution. It is then expected that other empirical data, perhaps
micromodeling, could be used to further validate the present methodology.

\section*{Declaration of competing interest}
The authors declare that they have no known competing financial interests
or personal relationships that could have appeared to influence the work
reported in this paper.

\section*{Acknowledgments}

B.J.S.\ thanks Amanda R. Lopes for her earlier advice on Python
programming, and also acknowledges the financial support from
\textit{Coordena{\c{c}\~{a}}o de Aperfei{\c{c}}oamento de Pessoal de
N\'{\i}vel Superior - Brasil (CAPES)} - Finance Code 001. M.B.R.\
received partial financial support from FAPERJ -- \textit{Rio de
Janeiro State Research Funding Agency}, grant number E-26/210.552/2024.













\bibliography{cosmo}
\bibliographystyle{elsarticle-num}
\end{document}